\documentclass[a4paper,11pt]{article}
\usepackage{jinstpub} % for details on the use of the package, please see the JINST-author-manual
\usepackage{lineno}
\title{Study of cosmogenic activation above ground of Ar for DarkSide-20k}

\author{S. Cebrián, on behalf of the DarkSide-20k Collaboration}
\affiliation{Centro de Astropartículas y Física de Altas Energías, Universidad de Zaragoza\\
Zaragoza 50009, Spain}
\emailAdd{ds-ed@lngs.infn.it}

\abstract{The production of long-lived radioactive isotopes due to the exposure to cosmic rays on the Earth's surface is an hazard for experiments searching for rare events like the direct detection of galactic dark matter particles. The use of large amounts of liquid argon is foreseen in different projects, like the DarkSide-20k experiment, intended to look for Weakly Interacting Massive Particles at the Laboratori Nazionali del Gran Sasso. Here, results from the study of the cosmogenic activation of argon carried out in the context of DarkSide-20k are presented. The induced activity of several isotopes, including $^{39}$Ar, and the expected counting rates in the detector have been deduced, considering exposure conditions as realistic as possible.}


\begin{document}
\maketitle
\flushbottom

\section{Introduction}
\label{sec:intro}

% Dark Matter, DarkSide-20k
To investigate the nature of the dark matter which is expected to fill our galaxy, one approach is to search for Weakly Interacting Massive Particles (WIMPs) by direct detection via WIMP-nucleus elastic scattering using different kinds of sensitive radiation detectors~\cite{appecdm}. Noble elements are ideal targets because the material is easily purified and detectors can be scaled in mass for high sensitivity; liquid argon (LAr) provides an outstanding capability to separate electron recoils (ER) from nuclear recoil (NR) events. The Global Argon Dark Matter Collaboration (GADMC) has been established to push the sensitivity for WIMP detection down through the neutrino fog. The first step will be the operation of the DarkSide-20k experiment at the Laboratori Nazionali del Gran Sasso (LNGS) in Italy. The core of the DarkSide-20k apparatus is a dual-phase Time Projection Chamber (TPC), serving as active WIMP target, filled by low-radioactivity underground argon (UAr) \cite{ricap}; a total of 99.2~t of UAr is required, 51.1~t inside the TPC and the rest in the neutron veto. It is planned to produce 120~t of UAr considering contingency. SiPMs in Photo-Detector Modules read the prompt scintillation in the liquid (S1) and delayed electroluminescence in the gas phase (S2). The TPC walls are made of a gadolinium-loaded acrylic vessel (Gd-PMMA) to tag neutron-induced background events. The detector is housed within a 12-ton vessel, made of stainless steel, immersed in a bath of 700~t of atmospheric argon (AAr) acting as radiation shield and outer veto detector for cosmic background. The experiment is designed with a goal of an instrumental background $<$0.1~events over a 200~t$\cdot$y exposure for a fiducial mass of 20~t and thus ultra-low background conditions are required. Despite the excellent background discrimination capabilities in DarkSide-20k, acceptance losses (via ER + NR pile-up in the TPC or accidental coincidence between the Veto and TPC signals that mimic the neutron capture signature) can be produced by $\gamma$ or $\beta$ emitters in the set-up and, therefore, these background sources must be carefully considered too.

%Activation
In the context of rare event searches, the long-lived radioactive isotopes induced in the materials of the experiments by the exposure to cosmic rays during fabrication, transport and storage have to be taken into account as an important background source \cite{cebrian,cebrianuniverse}. One of the most relevant processes in the cosmogenic activation is the spallation of nuclei by high energy nucleons but other reactions can be important too. It is desirable to have reliable estimates of activation yields to properly define the strategies to keep it under control by minimizing exposure or even using shielding against cosmic rays. Concerning argon, the relevant cosmogenically produced radioactive isotopes are $^{39}$Ar and also $^{37}$Ar, $^{42}$Ar and $^3$H.
\begin{itemize}
\item $^{39}$Ar is a $\beta^{-}$ emitter with a transition energy of 565~keV and half-life of 269~y \cite{toi}; it is mainly produced by the $^{40}$Ar(n,2n)$^{39}$Ar reaction by cosmic neutrons. The typical activity of $^{39}$Ar in AAr is at the level of $\sim$1~Bq/kg. The DarkSide-50 experiment at LNGS showed the feasibility of using UAr from deep CO$_2$ wells in Colorado; the measured activity of $^{39}$Ar was (0.73 $\pm$ 0.11)~mBq/kg, corresponding to a reduction factor relative to the AAr of 1400$\pm$200 \cite{darkside50}.
\item The presence of cosmogenically produced $^{37}$Ar was also detected at the beginning of the DarkSide-50 operation. It decays 100\% by electron capture to the ground state of the daughter nuclei with a half-life of 35.02~days \cite{ddep}; then, the binding energy of electrons from K-shell (2.8~keV, at 90.21\%) and L-shell (0.20-0.27~keV, at 8.72\%) gives a distinctive signature. The main production channel is the $^{40}$Ar(n,4n)$^{37}$Ar reaction. 
\item $^{42}$Ar is a pure $\beta^-$ emitter with a 32.9~y half-life and transition energy of 599~keV, generating $^{42}$K, also a $\beta^-$ emitter with half-life of 12.36~h and transition energy of 3525~keV \cite{toi}; this isotope can affect neutrinoless double beta experiments using liquid argon as cooling bath and shielding, as shown by the GERDA experiment \cite{gerda42ar} and its specific activity has been studied by ICARUS \cite{icarus}, DBA (92$^{+22}_{-46}$ $\mu$Bq/kg \cite{dba}) and DEAP (40.4$\pm$5.9~$\mu$Bq/kg \cite{deap}). Here, the effect of $^{42}$Ar in DarkSide-20k has not been evaluated.
\item $^3$H is a pure $\beta^-$ emitter with transition energy of 18.6~keV and a long half-life of 12.3~y~\cite{ddep}; therefore, it is particularly relevant for dark matter targets \cite{tritiumpaper}. In principle, purification systems for LAr may remove all non-argon radionuclides; activated $^3$H can be separated from argon with SAES Getters \cite{meikrantzTritiumProcessApplications1995} and will be removed {\it in situ} while the UAr recirculates.
%$^3$H should not be a problem for DarkSide-20k
\end{itemize}

%projects
For the procurement of large amounts of low-radioactivity UAr as detector target there are three projects underway in the GADMC: 
\begin{itemize}
\item Extraction of argon from an underground source will be carried out at the Urania plant, in Cortez, CO (US). This is the same source used for the DarkSide-50 detector.
\item UAr will be further chemically purified to detector-grade argon in the Aria facility, in Sardinia (Italy), to remove non-argon isotopes. Aria can also be operated in isotopic separation mode to achieve a 10-fold suppression of $^{39}$Ar although at a much reduced throughput \cite{aria,aria2}; this further suppression beyond UAr level is not needed to achieve the physics goals of DarkSide-20k. 
\item Assessing the ultra-low $^{39}$Ar content of the UAr is the the goal of the DArT detector \cite{dart} in construction at the Canfranc Underground Laboratory (LSC) in Spain. 
\end{itemize}

Activation of $^{39}$Ar and $^{37}$Ar in argon has been measured in \cite{saldanha} and quantified by GEANT4 simulation, considering also other radioisotopes, in \cite{zhangmei}. A specific study of the cosmogenic activation above ground for the DarkSide-20k experiment has been made in \cite{dsact} assuming exposure on the Earth's surface under realistic conditions in order to quantify the yields for all the relevant detector materials and the effect on the expected counting rates in comparison with that of other radioactive backgrounds. Here, after briefly describing the methodology of the study in section~\ref{sec:meth}, the main results related to argon are summarized in section~\ref{sec:res}, considering different conditions for the transport of the UAr from US to Italy.

\section{Methodology}
\label{sec:meth}

%When possible, align equations on the equal sign. The package \texttt{amsmath} is already loaded. See \eqref{eq:x}.
%\begin{equation}
%\label{eq:x}
%\begin{aligned}
%x &= 1 \,,
%\qquad
%y = 2 \,,
%\\
%z &= 3 \,.
%\end{aligned}
%\end{equation}
%Also, watch out for the punctuation at the end of the equations.

%general
To quantify the effect of material cosmogenic activation in a particular experiment, the first step is to know the production rates, $R$, of the relevant isotopes induced in the material targets. Then, the produced activity, $A$, can be estimated according to the exposure history to cosmic rays; for instance, considering just a time of exposure $t_{exp}$ followed by a cooling time (time spent underground once shielded from cosmic rays) $t_{cool}$, for an isotope with decay constant $\lambda$, the activity can be evaluated as:
\begin{equation}
\label{eqact}
\begin{aligned}
A = R [1-\exp(-\lambda t_{exp})] \exp(-\lambda t_{cool}). 
\end{aligned}
\end{equation}
\noindent Finally, the counting rate generated in the detector by this activity can be computed by Monte Carlo simulation. Direct measurements of production rates at sea level have been carried out for a few materials from the saturation activity, obtained by sensitive screening of samples exposed in well-controlled conditions or by irradiating samples in high flux particle beams. However, in many cases, production rates must be evaluated from the flux of cosmic rays, $\phi$, and the isotope production cross-section, $\sigma$, with both dependent on the particle energy $E$:
\begin{equation}
\label{eqrate}
\begin{aligned}
R=N_t\int\sigma(E)\phi(E)dE,
\end{aligned}
\end{equation}
\noindent where $N_t$ is the number of target nuclei. The spread for different calculations of productions rates is usually important, even within a factor 2.

%DarkSide-20K
For DarkSide-20k estimates, production rates at sea level from \cite{saldanha} have been used for $^{39}$Ar and $^{37}$Ar, including results from controlled irradiation at Los Alamos Neutron Science Center (LANSCE) with a neutron beam resembling the cosmic neutron spectrum and the quantification of other production mechanisms due to muon capture, cosmic protons and high energy $\gamma$ rays. For $^{3}$H, a dedicated calculation of the production rate has been performed giving  (168$\pm$53)~atoms/kg/day \cite{dsact}. For these calculations, the analytic expression for the cosmic neutron spectrum at sea level presented by Gordon et al \cite{gordon}, deduced by fitting data from a set of measurements for energies above 0.4~MeV, has been used; for the cross sections, different codes, databases and libraries from Monte Carlo simulation have been considered: the Experimental Nuclear Reaction Data database (EXFOR, CSISRS in US)~\cite{exfor}; the YIELDX routine based on the semiempirical Silberberg and Tsao equations~\cite{tsao1,tsao2,tsao3};  TENDL (TALYS-based Evaluated Nuclear Data Library) \cite{tendl}, based on the TALYS code, for protons and neutrons with energies up to 200~MeV; JENDL (Japanese Evaluated Nuclear Data Library) \cite{jendl} High Energy File, based on the GNASH code, for protons and neutrons from 20~MeV to 3~GeV; and HEAD-2009 (High Energy Activation Data) \cite{head2009} for protons and neutrons with higher energies, from 150~MeV up to 1~GeV based on a selection of models and codes.

%correction factors

The UAr to be used in DarkSide-20k is extracted at the Urania facility, in Colorado, at a quite high altitude, so a correction factor $f$ to the cosmic ray flux at sea level must be taken into consideration. As described in~\cite{ziegler}, the intensities $I_1$ and $I_2$ at two different altitudes $A_1$ and $A_2$ are related as:
\begin{equation}
\label{eqI}
\begin{aligned}
    I_2=I_1 \exp[(A_1-A_2)/L], 
\end{aligned}
\end{equation}
\noindent being $L$ the absorption length for the cosmic ray particles. For neutrons, reported values of $f$ at Colorado locations have been adjusted to the altitude at the Urania facilities (at 2164~m) obtaining $f=6.43$ \cite{dsact}. For cosmic protons and muons, the correction factors have been obtained just from Eq.~\ref{eqI} considering the corresponding absorption lengths \cite{ziegler}, as $f=8.67$ for protons and $f=2.48$ for muons.

\section{Results}
\label{sec:res}

For DarkSide-20k, the UAr extracted at the Urania plant will be shipped firstly to the Aria facility for purification and then to LNGS for storage and final operation. 

\subsection{First estimate of activities and counting rates}
%article: gas transport
A total of 120~t of UAr will be produced for DarkSide-20k. The initial plan was to ship the UAr in high-pressure gas cylinders organized into skids capable of containing $\sim$2~t of UAr each, repeating the whole shipment procedure for arrays of three skids (6~t). Since the time estimated to fill one skid is 8~days, the first skid would have an exposure of 24~days while being stored at Urania, before starting transportation, the second skid 16~days and the third skid 8~days. Afterwards, the foreseen exposure times were: 7~days for the trip from Urania to a shipping port; 67~days for the trip overseas to Sardinia; 60~days to process each of the two batches of 60~t each for purification of UAr at Aria (the start of the processing at Aria requires the arrival of 60~t of UAr); and 10~days for the trip from Aria to LNGS. Table \ref{tableSummary} summarizes the induced activity estimated in these conditions of the analyzed cosmogenic products as well as the corresponding expected counting rates in the TPC and inner veto when all the UAr arrives to LNGS, as presented in \cite{dsact}; the relative contribution of each exposure step can be found in Tables~6 and 7 of this reference. For $^{39}$Ar, the induced activity was just 2.8\% of the residual activity measured in DarkSide-50 for UAr of the same source and thus considered tolerable. It must be noted that assuming the activities from an extensive material screening campaign, preliminary estimates of $\gamma$ background rates point to values around 50~Hz in the TPC and 100~Hz in the neutron Veto; the $\beta$ contribution of $^{39}$Ar, from the measured activity of $^{39}$Ar in DarkSide-50, yields 36~Hz in the TPC and 26~Hz in the Veto.
%The quantified effect of some uncertain steps in the procedure of UAr production shows that there is enough contingency.

%\begin{table}[htbp]
%\centering
%\caption{We prefer to have top and bottom borders around the tables.\label{tab:i}}
%\smallskip
%\begin{tabular}{lr|c}
%\hline
%x&y&x and y\\
%\hline
%a & b & a and b\\
%1 & 2 & 1 and 2\\
%$\alpha$ & $\beta$ & $\alpha$ and $\beta$\\
%\hline
%\end{tabular}
%\end{table}

\begin{table}[htbp]
\centering
\caption{Cosmogenically induced activity and counting rates in DarkSide-20k TPC and inner veto estimated in the exposure conditions assuming transport of gaseous UAr, as in \cite{dsact}. All reported values correspond to the moment when the materials are brought underground. For $^{3}$H, row (1) and (2) assume no purification and ideal purification at Aria, respectively.}
\label{tableSummary}
\smallskip
\begin{tabular}{lccc}
\hline
Isotope  & Activity  & TPC rate  & Veto rate  \\
&  ($\mu$Bq/kg) & (Hz) & (Hz) \\ \hline
$^{39}$Ar & 20.7$\pm$2.8 & 1.03$\pm$0.14 & 0.662$\pm$0.090 \\ 
$^{37}$Ar & 103$\pm$14 & 5.15$\pm$0.68 &  3.30$\pm$0.43 \\ 
$^{3}$H (1) & 76$\pm$24 & 3.8$\pm$1.2& 2.42$\pm$0.76 \\
$^{3}$H (2) & 2.97$\pm$0.94 & 0.148$\pm$0.047 & 0.095$\pm$0.030 \\ \hline 	
\end{tabular}
\end{table}	

\subsection{Impact on $^{39}$Ar activity of alternative shipment and processing schemes}

%per month
As it could be necessary to modify the conditions of the shipment and processing of the UAr, it is important to consider the impact on the induced $^{39}$Ar activity of different exposure times. The activity produced after one month of exposure is evaluated to be (13.3$\pm$2.0)~$\mu$Bq/kg at Urania and (2.57$\pm$0.32)~$\mu$Bq/kg at sea level; as the $^{39}$Ar half-life is very large, these values allow to estimate for different times the almost linear increase of activity given by Eq.~\ref{eqact}. Concerning the counting rates in the detector, they can be easily evaluated taking into account that one $\mu$Bq/kg of $^{39}$Ar produces 0.050~Hz in the TPC and 0.032~Hz in the veto.

%new: liquid transport
Presently, it is being considered to transport UAr in liquid phase and then the organization of transportation is different, which changes significantly the exposure times at Urania and Aria facilities. An update of the results of the cosmogenic activation study for DarkSide-20k is ongoing. As a first estimate of the effect of the new exposure conditions, the induced activity of $^{39}$Ar has been re-evaluated following the same methodology but considering 60 days of exposure at Urania and 6 months of exposure at Aria; the conditions of the transport in the US and overseas are the same ones described before. Then, at the end of Aria processing the activity would be (49.7$\pm$7.0)~$\mu$Bq/kg, being 6.8\% of the measured DarkSide-50 activity. The relative contribution to the final yield is 54\% for Urania and 31\% for Aria.  

%Considering as an upper limit 90 days (instead of 60 days) of exposure at Urania, the induced activity of $^{39}$Ar would be (63.0$\pm$9.0)~$\mu$Bq/kg, 8.6\% of the  DarkSide-50 activity.
%giving a counting rate of (2.48$\pm$0.35)~Hz in the TPC and (1.59$\pm$0.22)~Hz in the veto
 
\section{Conclusions}

For DarkSide-20k, cosmogenic activation as a source of $\beta/\gamma$ background is being quantified for LAr from realistic exposure conditions in order to assess the contribution to the counting rates and decide if additional exposure restrictions are necessary. A total of 120~t of UAr depleted in $^{39}$Ar must be extracted and processed for filling the TPC and inner veto and induced cosmogenic activity must not significantly increase the expected very low activity level of $^{39}$Ar. The possible induced activity on surface, from the extraction at Urania to the storage at LNGS, has been analyzed not only for $^{39}$Ar but also for $^{37}$Ar and $^{3}$H. Production rates from Ref.~\cite{saldanha}, based on a neutron irradiation experiment, have been considered for the Ar isotopes while for $^{3}$H an estimate of the production rate by cosmic neutrons has been used. Contributions from $^{37}$Ar and $^{3}$H are not problematic thanks to short half-life and purification, respectively. Considering a detailed exposure history assuming transport of gaseous UAr, the induced activity of $^{39}$Ar was just 2.8\% of the residual activity measured in DarkSide-50 for UAr of the same source \cite{dsact}. A re-evaluation of the cosmogenic activation is ongoing as the change to transport UAr in liquid phase makes necessary reconsidering the exposure times at Urania and Aria facilities; assuming 60 days and 6 months, respectively, $^{39}$Ar would be increased by a factor $\sim$2.5, being at the level of 50~$\mu$Bq/kg, still well below the uncertainty in the expected $^{39}$Ar activity from DarkSide-50.

There is a growing interest in the use of ultra-pure UAr outside GADMC \cite{pnnlworkshop,SnowmassUArFacility}; the results of this study could be useful to set exposure limitations in future LAr projects related not only with dark matter searches but also with neutrino studies.

\end{document}